\documentclass[11pt,a4paper]{article}
\usepackage{latexsym}
\usepackage{graphicx}
\usepackage{epsfig}
\usepackage{amssymb}
\usepackage{amsmath}
\usepackage{epstopdf}
\usepackage{eufrak}
\usepackage{epstopdf}
\usepackage{pb-diagram}
\usepackage{url}

\usepackage[hmargin=2.5cm, vmargin=2.5cm]{geometry} 

\newcommand\Beq{\begin{eqnarray}} 
\newcommand\Eeq{\end{eqnarray}}

\newcommand\Bfig{\begin{figure}} 
\newcommand\Efig{\end{figure}}

\newcommand{\eq}[1]{(\ref{#1})}
\newcommand{\eqs}[2]{(\ref{#1})~\&~(\ref{#2})}
\newcommand{\eqss}[2]{(\ref{#1})--(\ref{#2})}

\renewcommand{\sec}[1]{\S\ref{#1}}

\newcommand{\Eq}[1]{(\ref{#1})}
\newcommand{\Eqs}[2]{(\ref{#1})~\&~(\ref{#2})}

\newcommand{\dd}{\,\mathrm{d}}
\newcommand{\h}{d}
\renewcommand{\H}{L}

\newcommand{\pd}[1]{\partial_{#1}}

\newcommand{\eps}{\varepsilon}
\newcommand{\f}{f}
\newcommand{\Ro}{q}

\renewcommand{\vec}[1]{\boldsymbol{#1}}

\newcommand{\n}{\\ }

\newcommand{\paper}[8]{ \bibitem{#1} {#2} {#3} {#4}. \textit{#5} \textbf{#6}, {#7}--{#8}. }
\newcommand{\book}[7]{ \bibitem{#1} {#2} {#3} \textit{#4}. {#5}: {#6} }

\begin{document}

\title{On the magnetorotational instability and elastic buckling}

\author{
Geoffrey M. Vasil$^{1}$}

\centerline{\Large \bf On the magnetorotational instability and elastic buckling}
\bigskip
\centerline{\large{Geoff Vasil} : \url{geoffrey.vasil@sydney.edu.au}}
\medskip
\centerline{\textsl{{University of Sydney School of Mathematics and Statistics, Sydney, NSW 2006, Australia}}}

\renewcommand*{\thefootnote}{\arabic{footnote}}

\bigskip
\centerline{\large 10 March 2015}
\bigskip

\bigskip
\textbf{Abstract}

This paper demonstrates an equivalence between rotating magnetised shear flows and a stressed elastic beam. This results from finding the same form of dynamical equations after an asymptotic reduction of the axis-symmetric magnetorotational instability (MRI) under the assumption of almost-critical driving.  The analysis considers the MRI dynamics in a non-dissipative near-equilibrium regime. Both the magnetic and elastic systems reduce to a simple one-dimensional wave equation with a nonlocal nonlinear feedback.   Under transformation, the equation comprises a large number of mean-field interacting Duffing oscillators.  This system was the first proven example of a strange attractor in a partial differential equation.   Finding the same reduced equation in two natural applications suggests the model might result from other applications and could fall into a universal class based on symmetry.


\section{Introduction \label{sec:intro}}

In a non-dissipative, purely hydrodynamical context, the 
famous Rayleigh criterion \cite{rayleigh,drazin_reid} of outwardly increasing angular 
momentum governs the stability of rotating shear flows. That is,
\Beq
\label{Rayleigh-condition}
\frac{d}{dr}\left( r^{4} \Omega(r)^{2}\right) < 0.
\Eeq
gives a necessary condition for the axis-symmetric instability of a given radially varying cylindrical rotation profile, where $r$ represents the outward 
directed radial coordinate, and $\Omega(r)$ represents the local 
angular rotation rate.  During the 1950s and 1960s, Chandrasekhar \cite{chandra_a1,chandra_a2} in The West, and Velikov \cite{velikov} in the former Soviet Union considered the effect of an axial magnetic field on the stability of a cylindrically swirling conducting fluid, thereby modifying the  Rayleigh criterion.  In these works, both authors found independently that (in the ideal non-dissipative 
regime), the presence of a magnetic field can catalyse an instability 
for a merely inwardly increasing angular frequency profile; rather than 
angular momentum.  That is, a magnetohydrodynamic (MHD) instability only requires 
\Beq
\frac{d}{dr}\left(\Omega(r)^{2} \right) < 0,
\Eeq
independent of the strength of the axial magnetic field (in a 
sufficiently large domain).  However, in spite of its discussion in 
Chandrasekhar's famous book on fluid instabilities \cite{chandra_b}, this result 
went mostly unnoticed until its application in astrophysics.  

In the early 1990s, Balbus and Hawley auspiciously applied the 
destabilising of nature of a weak magnetic field to solve a previous 
astrophysical paradox \cite{B&H91}.  Specifically, in a fluid system orbiting a 
central mass, $M$, a balance between gravitational and centrifugal 
acceleration chiefly determines the rotation profile, $r \Omega(r)^{2} = GM/r^{2}$; \ G represents Newton's gravitational constant.  This \textit{Keplerian} profile implies 
\Beq
\frac{d}{dr}\left( r^{4} \Omega(r)^{2}\right) = G M > 0,
\Eeq
which ostensibly guarantees stability to infinitesimal perturbations, according to the Rayleigh criterion.
However, the stability of Keplerian differential rotation does not reconcile with observations 
of large brightnesses in proto-planetary disks \cite{B&H98}.  Balbus 
and Hawley invoked the presence of a weak background magnetic 
field and employed the more favourable instability requirement
\Beq
\frac{d}{dr}\left( \Omega(r)^{2}\right) = - 3 \frac{\Omega(r)^{2}}{r} < 0.
\Eeq
This argument, along with their nonlinear numerical simulations, provided very strong evidence supporting the hypothesis of magnetised accretion disk dynamics \cite{H&B91,H&B92}.  

The advent of the magnetorotational instability (MRI) in astrophysics produced many notable results over the last three 
decades \cite{B&H98}.  Little doubt exists regarding the operation and efficiency of the MRI in hot astrophysical disk systems.  But questions do remain concerning the instability's cessation and saturation \cite{goodman_xu,julien_knobloch}.  This paper explores the dynamics of the MRI in a controlled setting. This simple approach helps explain some mysterious aspects regarding how the instability transports momentum and magnetic flux, which proves useful in understanding possible saturation mechanisms. 

In contrast to a magnetised fluid, an elastic material is a continuum substance that returns to its original form with the removal of applied loads.  In spite of clear differences, elastodynamics and MHD contain many deep physical and mathematical similarities. And just with forced fluids, elastic materials possess an extremely rich range of behaviours, specifically including dynamical instabilities.  

When a solid column \textit{buckles}, it looses the ability to support a load while retaining its elastic integrity.  Because of its obvious importance to the stability of structures, the scientific investigation of buckling dates well back into antiquity, and quantitative investigations started at the very beginning of the modern scientific era.  As early as the 1480s, Leonardo da Vinci produced empirical criteria addressing the stability and lateral deflection of columns under compression \cite{godoy}.  In the 18th century, Euler and Bernoulli began to consider time-dependence of elastic deformations and buckling \cite{johnston}. In particular, Euler derived a practical formula for calculating the critical load on a slender beam. Engineers still use this formula (and its extensions) in modern construction design \cite{domokos_etal}.  

Elastic substances exhibit many more types of instabilities than simple buckling \cite{bigoni}. After much success with linear dynamics of elastic solids, work in the 1970s began to focus heavily on nonlinear finite deformations. Aeronautical engineering and manufacturing particularly requires understanding the time-dependent effects of solid materials under a wide range of situations.  The introduction of methods from mathematical dynamical systems theory allowed rigorous analysis for a wide range of scenarios \cite{marsden_hughes}.  In a series of work, Holmes and Marsden considered a class of nonlinear, nonlocal models for buckling and fluttering \cite{holmes,holmes_marsden}.  They proved the first example in a partial differential equation of chaos and a strange attractor; \textit{i.e.}, an infinity of arbitrarily long periodic orbits. 

The current paper finds an almost identical equation arising naturally from the weakly nonlinear theory of the MRI in a simple geometry. This link gives qualitative insight into the astrophysical setting, and a more quantitative understanding of magneto-Taylor-Couette flow in laboratory experiments \cite{sisan_etal}.  In contrast to astrophysical disks, the model may also apply to the inside of some stars with stronger ambient magnetic field and more modest differential rotation.  However, even if no astrophysical object exist near-equilibrium, understanding nonlinear behaviour near a phase transition can uncover fundamental interactions. In this case, the current model helps elucidate transport of momentum and magnetic flux in an important MHD instability. To use a zoological analogy: underneath many obvious differences, both the buckling beam and the wild-type MRI contain a morphologically equivalent primitive skeleton.  
 
The summary of the remained of this paper follows: \S2 analyses both the linear and nonlinear theory of the MRI. The main result of the paper follows from \eqs{WNLMRI}{reduced-sigma-hat} and its analogy to \eq{BEAM}; \S3 derives the buckling instability of a slender elastic beam; \S4 provides a detailed discussion on several aspects of the MRI and the buckling beam; \S5 shows computational results of the nonlinear solutions; \S6 provides conclusions. 
    
\Bfig
\begin{center}
\includegraphics[scale=0.24]{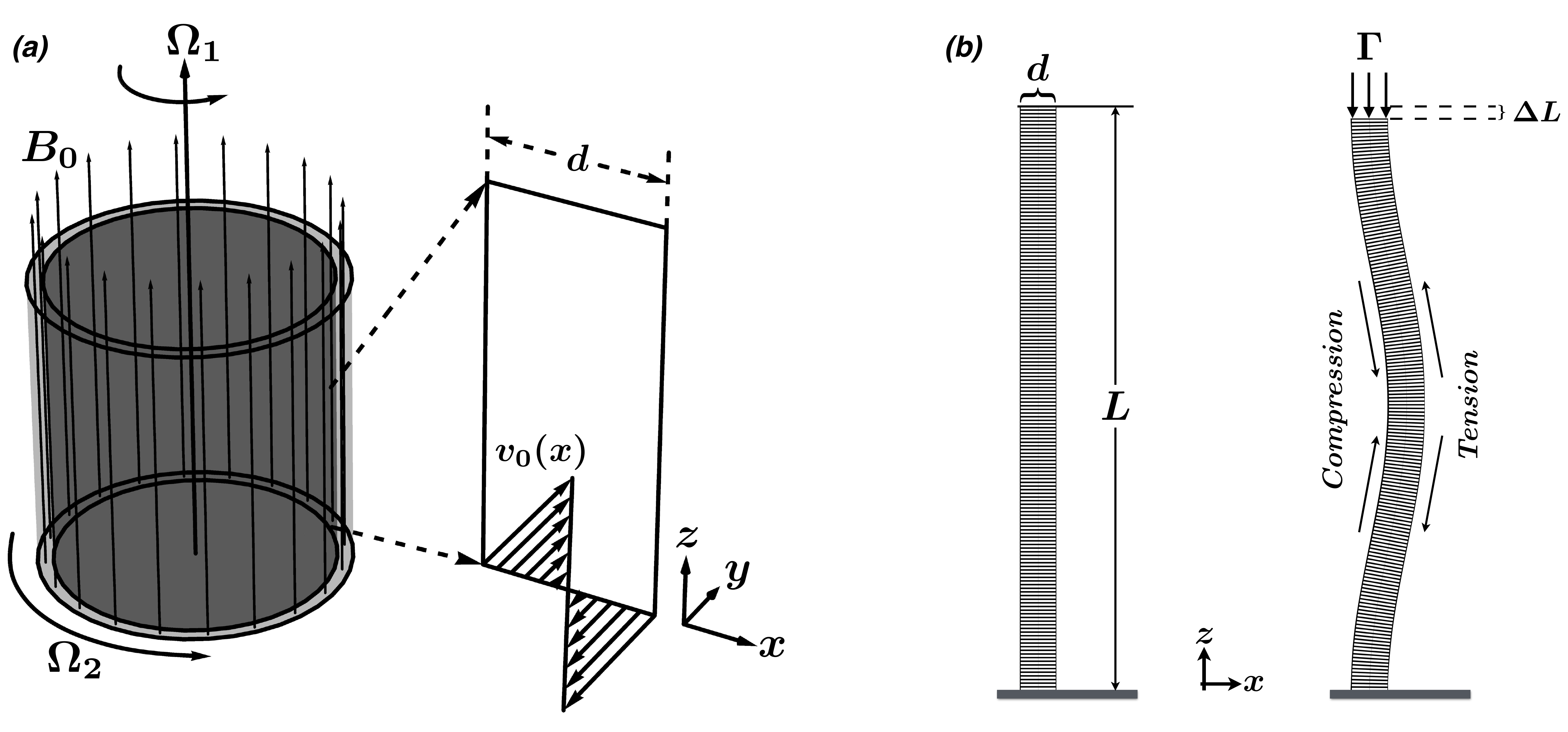}
\caption{a: Fluid system digram. The model considered in \S\ref{sec:MHD} assumes a narrow-gap Cartesian limit of an axis-symmetric Taylor-Couette cylinder. Different rotation rates on the inner and outer cylinders manifests as a local shear in the rotating frame. b: Elastic system diagram. The left-side image depicts an un-stressed beam of length $L$, and thickness $\h$.  The right-side image depicts the solution to the slightly buckled beam with clamped-end boundary conditions. 
\label{diagram}}
\end{center}
\Efig

\section{Instability of a rotating MHD shear flow \label{sec:MHD}}

The first part of this paper considers a two-dimensional model for the 
MRI of an incompressible magnetised fluid with linear shear flow.  
The system remains invariant along the shear 
direction, but contains all three components of flow and magnetic 
field.  The full primitive equations describe the dynamics in a rotating coordinate system, 
\Beq
\label{vector-v}
\pd{t}\vec{v} + \vec{v}\cdot \vec{\nabla} \vec{v} + 2 \Omega \vec{\hat{z}}\vec{\times}\vec{v} +\vec{\nabla} p   &=&   \vec{B}\cdot \vec{\nabla} \vec{B} +  \nu \nabla^{2} \vec{v} \n
\label{div-u} \vec{\nabla}\cdot\vec{v} &=& 0 \n
\label{vector-B} \pd{t}\vec{B} + \vec{v}\cdot \vec{\nabla} \vec{B} &=& \vec{B}\cdot \vec{\nabla} \vec{v} + \eta \nabla^{2} \vec{B}\n
\label{div-B} \vec{\nabla}\cdot\vec{B} &=& 0.
\Eeq
\eqss{vector-v}{div-B} govern the dynamics of all three components of flow and magnetic field, but make the implicit assumption of axis-symmetry $\pd{y} = 0$.

The stream-function, and magnetic-scalar-potential 
formulations enforce the solenoidal character of flow and 
magnetic field \eqs{div-u}{div-B},
\Beq
\label{u-def}
\vec{v} &=& \left[ v_{0}(x) + v(x,z) \right]\vec{\hat{y}} \ \ - \ \  
\vec{\hat{y}} \times \vec{\nabla} \psi(x,z) \n
\label{B-def}
\vec{B} &=& b(x,z)\,\vec{\hat{y}} \ \ -  \ \  \vec{\hat{y}} \times 
\vec{\nabla}  \left[ a_{0}(x) + a(x,z)\,\right].
\Eeq
In \eqs{u-def}{B-def}, $v(x,z)$, $b(x,z)$ represent the flow and 
magnetic field in the stream-wise direction respectively, and $
\psi(y,z)$, $a(y,z)$ represent the poloidal stream function and scalar 
potential in the plane perpendicular to the streaming flow.  Both 
$v(x,z)$ and $a(x,z)$ represent perturbations relative to the linear 
background profiles
\Beq
\label{v_0, a_0}
v_{0}(x) = S x-\frac{S\h}{2}, \quad a_{0}(x) = B x-\frac{B\h}{2},
\Eeq
The constant terms in \eq{v_0, a_0} ensures the mean of both 
quantities vanishes over the domain $0 \le x \le \h$.  This represents an arbitrary gauge for 
the magnetic potential, but complies with the definition of the 
global rotations rate for the shear flow.
The linear background magnetic potential implies a uniform 
background magnetic field in the vertical direction, $B \vec{\hat{z}}$.  
The parameter $S$ represents the local background shear rate. Figure \ref{diagram}a depicts the basic geometry and background configuration. 

Obtaining equations for the stream function and magnetic scalar potential requires taking the two-dimensional curl of the $x$ and $z$ components of the flow and magnetic filed equations. 
The incompressible magnetohydrodynamic equations for each 
scalar variable are  
\Beq
\label{b-eq} &&\pd{t}b + J(\psi,b) \ = \ J(a,v) + B\pd{z}v - S \pd{z}a + 
\eta \nabla^{2} b \n
\label{a-eq} &&\pd{t}a + J(\psi,a) \ = \  B\pd{z}\psi + \eta \nabla^{2} a
\n
\label{v-eq} && \pd{t}v + J(\psi,v) - (\f + S) \pd{z}\psi \ = \ J(a,b) + B
\pd{z}b + \nu \nabla^{2} u\n
\label{psi-eq} && \pd{t} \nabla^{2}\psi + J(\psi,\nabla^{2}\psi) +\f 
\pd{z}v\ = \ J(a,\nabla^{2}a) + B\pd{z}\nabla^{2}a + \nu \nabla^{4}\psi,
\Eeq
where
\Beq
J(p,q) \equiv \pd{x}p\pd{z}q - \pd{z}p \pd{x}q,
\quad
\nabla^{2} \equiv \pd{x}^{2} + \pd{z}^{2}
\Eeq
represent the nonlinear transport of two quantities, and  two-dimensional Laplacian respectively. For simplicity, we 
measure the magnetic field in Alfv\'{e}n units, \textit{i.e.},  $
\mu_{0}\rho_{0} = 1$. We restore these parameters at the 
conclusion of the derivations.  For the remaining parameters, $\f=2 
\Omega$ represents the background vorticity owing to the frame 
rotation, $\Omega$.  The dissipation parameters, $\nu$ and $\eta$ 
represents viscosity and magnetic diffusivity respectively.  The 
following analysis considers the cases of both small and/or 
completely vanishing diffusion coefficients.  

\subsection{Linear Analysis \label{linear-theory}}

Temporarily neglecting the nonlinear and dissipative terms in 
\eqss{b-eq}{psi-eq}  helps determine the relevant spatiotemporal scales.  After linearising, and with some simplifications,   
\Beq
\label{linear-b-eq} &&\left(\pd{t}^{2} - B^{2}\pd{z}^{2}\right)b \ = \  \f B 
\pd{z}^{2} \psi \n
\label{linear-a-eq} &&\pd{t}a  \ = \  B\pd{z}\psi \n
\label{linear-u-eq} && \left(\pd{t}^{2} - B^{2}\pd{z}^{2}\right)\pd{t}v  \ = 
\ \left( (\f + S)\pd{t}^{2} - S B^{2} \pd{z}^{2} \right)\pd{z}\psi\n
\label{linear-psi-eq} && \left(\pd{t}^{2} - B^{2}\pd{z}
^{2}\right)^{2}\nabla^{2}\psi +  \f\left( (\f + S)\pd{t}^{2} - S B^{2} \pd{z}
^{2} \right)\pd{z}^{2}\psi \ = \ 0.
\Eeq
For a given $\psi(t,x,z)$, \eqss{linear-b-eq}{linear-u-eq} determine $b(t,x,z)$, $a(t,x,z)$, and 
$v(t,x,z)$ respectively.  \Eq{linear-psi-eq} determines the stream function, subject to the impenetrable boundary conditions
\Beq
\psi|_{x=0} \ = \ \psi|_{x=\h} \ = \ 0.
\Eeq
Wave-like solutions 
\Beq
\psi = \Psi e^{i ( k z + \omega t )} \sin(\pi x /\h) + \mathrm{c.c.}
\Eeq
produce the ``dispersion relation'' for the (possibly complex-valued) 
frequency, $\omega$, versus wavenumber $k$, and other 
parameters, 
\Beq
\label{full dispersion}
\left(k^{2} + \frac{\pi^2}{\h^2} \right) \lambda^2  -  \f(\f+S)k^2 \lambda 
- \f^2 B^2  k^4  \ = \ 0,
\Eeq
where
\Beq
\lambda \equiv \omega^2 - B^2k^2.
\Eeq

\eq{full dispersion} gives purely real solutions for $\lambda$.  
Therefore, a transition from purely real frequencies ($\omega^2 > 0 $) 
to purely imaginary frequencies ($\omega^2 < 0 $) characterises 
any transition of stability. For any real $k$, 
\Beq
&& B^{2}\left(k^{2} + \frac{\pi^2}{\h^2} \right) +  \f S  \ = \ 0. 
\Eeq
gives the boundary ($\omega^{2} = 0$) between these two regimes.  Instability ($\omega^2 < 0 $) exists for
\Beq
 - \f S \ge B^{2}\left(k^{2} + \frac{\pi
   ^2}{\h^2}\right) > \frac{\pi^{2} B^{2}}{\h^{2}}.
\Eeq
For a finitely thick layer, 
\Beq
\label{S-crit}
S_{\mathrm{crit.}} \ \equiv \ -\frac{\pi^{2} B^{2}}{\f \h^{2}}.
\Eeq
defines the critical shear needed to drive the MRI.  

The difference between the critical shear and actual shear helps simplify the analysis in the vicinity of the instability, 
\Beq
\sigma  \equiv S_{\mathrm{crit.}} - S.
\Eeq
Assuming long-wavelength perturbations $|k \h| \ll 1$, and near-critical shear $|\sigma| \ll |S_{\mathrm{crit.}}|$, we can expand the 
dispersion relation to lowest order and find  
\Beq 
\label{approx-omega}
\omega^2 \approx 
\frac{B^2 \h^2 k^2 \left(B^2
   k^2-\f \sigma \right)}{\pi^2 B^2+\f^2 \h^2}.
\Eeq
\Eq{approx-omega} shows that $\omega =  \mathcal{O}(\sigma)$, and $k = \mathcal{O}(\sqrt{|\sigma|})$ for $\f |\sigma| \h^{2}/B^{2} \ll 1$. Figure \ref{stability regions}a shows the behaviour of \eq{approx-omega}.

Reintroducing space and time variables via $\omega \to - i \pd{t}$ and $k \to - i  \pd{z}$ implies  (heuristically) that \eq{approx-omega} represents the simplified (unstable) wave equation
\Beq 
\label{approx-wave}
\pd{t}^2\psi \approx 
-\frac{B^2 \h^2 }{\pi^2 B^2+\f^2 \h^2} \left(B^2
   \pd{z}^2+\f \sigma \right)\pd{z}^2\psi + \textit{Nonlinear Terms}.
\Eeq
\Eq{approx-wave} implicitly contains missing nonlinearities required to saturate exponential growth.   
The next section determines these unknown terms.

\Bfig
\begin{center}
\includegraphics[scale=0.19]{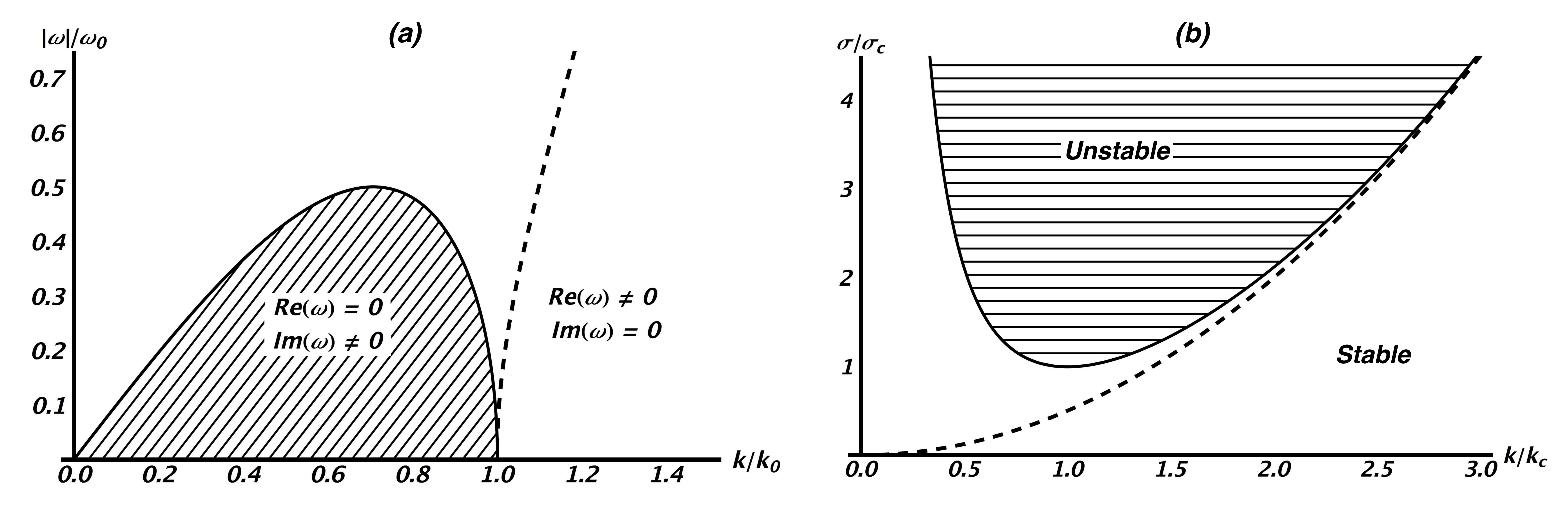}
\caption{(a): A ``dispersion curve'' showing the complex-valued 
frequency as a function of vertical wavenumber. The shaded grey 
region shows purely imaginary values, and the white region to the 
right of the dashed line corresponds to purely real waves. Assuming 
$\sigma > 0$, $k_{0} \equiv \sqrt{f \sigma|}/|B|$, $\omega_{0} \equiv 
f \sigma \h/\sqrt{\pi^{2}B^{2} + f^{2}\h^{2}}$.  The maximum growth 
rate occurs for $k = k_{0}/\sqrt{2}$ and $\omega = \omega_{0}/2$. (b): The solid line shows the critical stability curve for $\sigma(k)$ with finite dissipation.  The dashed line shows the corresponding curve with no dissipation. See \eqss{dissipation-sigma(k)}{Re_M,c-eq} for details.
\label{stability regions}}
\end{center}
\Efig

\subsection{Nonlinear Asymptotic Analysis}

Rendering the dynamical equations in dimensionless form often helps to simplify derivations.  In this case, working with scaled quantities 
partially obscures the physical meaning of different dynamical 
ingredients.  Therefore, the following analysis introduces a small 
non-dimensional parameter ($\eps \ll 1$) only to keep track of the relative 
magnitudes of the different terms.  We rescale all relevant dynamical 
variables in terms of particular powers of $\eps$ and conduct an 
asymptotic analysis accordingly.   From \eq{approx-omega} we deduce the consistent rescaling 
\Beq
&& S \to S_{0} - \eps^{2}\sigma, \quad \pd{t} \to \eps^{2} \pd{t},\quad 
\pd{x} \to \pd{x},\quad \pd{z} \to \eps\pd{z}
\Eeq
The linear system \eqss{linear-b-eq}{linear-u-eq} provides the relative 
amplitudes of $b$, $a$ and $v$ in terms of $\psi$.  The fully nonlinear dissipative system, \eqss{b-eq}
{psi-eq}, constrains the amplitude of $\psi$, and also gives the 
magnitude of $\nu$ and $\eta$ that allow dynamically significant 
dissipation.  That is,
\Beq
&&b \to \eps^{2} b, \quad a\to \eps a,\quad  v \to \eps v ,\quad   \psi 
\to \eps^{2} \psi  \n
&& \eta \to  \eps^{2}\eta ,\quad \nu \to  \eps^{2}\nu
\Eeq

Upon rescaling, \eqss{b-eq}{psi-eq} become 
\Beq
\label{rescaled-v-eq}
&& \pd{z}\left(B v - S_{0} a\right) \ = \ -\eps J(a,v)  + 
\eps^{2}\left(\pd{t}b - \sigma\pd{z}a - \eta \pd{x}^{2}b \right) + 
\mathcal{O}(\eps^{3})
\n 
&& \label{rescaled-a-eq} \pd{z}\left(B\pd{x}^{2}a - \f v\right) = - \eps 
J(a,\pd{x}^{2}a) + \eps^{2}\left( \pd{t} \pd{x}^{2}\psi  - B\pd{z}^{3}a - 
\nu \pd{x}^{4}\psi   \right) + \mathcal{O}(\eps^{3})\quad \quad 
\n
&&\label{rescaled-B-eq} \pd{t}a -B\pd{z}\psi - \eta \pd{x}^{2} a\ = \   - 
\eps  J(\psi,a) \n
&& \label{rescaled-psi-eq}  \pd{t}v - B\pd{z}b  - (\f + S_{0}) \pd{z}\psi 
- \nu \pd{x}^{2} v =  \eps \left(  J(a,b) -  J(\psi,v) \right) + \mathcal{O}
(\eps^{2}) 
\Eeq 
Considering this system order-by-order, the analysis halts when the 
first time derivative from the right-hand side enters the balance.  
Therefore, we may automatically neglect terms that are formally 
smaller than the highest order time-derivative. Other than not writing higher-order terms, \eqss{rescaled-v-eq}{rescaled-psi-eq} are completely equivalent to \eqss{b-eq}{psi-eq}.

We expand all dynamical variables in the following series 
\Beq
&&v = v_{0} + \eps v_{1} + \eps^{2} v_{2}  + \ldots, \quad 
a = a_{0} + \eps a_{1} + \eps^{2} a_{2} + \ldots\n
&&b = b_{0} + \eps b_{1} + \eps^{2} b_{2} + \ldots, \quad
\psi = \psi_{0} + \eps \psi_{1} + \eps^{2} \psi_{2}  + \ldots,
\Eeq
substitute this into \eqss{rescaled-v-eq}{rescaled-psi-eq}, and collect 
terms order-by-order in powers of $\eps$. The leading-order balance 
produces 
\Beq
\label{leading-order-1}
&& \pd{z}\left(B v_{0} - S_{0} a_{0}\right) \ = \ 0, \quad \quad
 \pd{z}\left(B\pd{x}^{2}a_{0} - \f v_{0}\right) = 0
\n
\label{leading-order-4}
&&\pd{t}a_{0} -B\pd{z}\psi_{0} \ = \  \eta \pd{x}^{2} a_{0} ,\quad \pd{t}v_{0} - B\pd{z}b_{0}  - (\f + S_{0}) \pd{z}\psi_{0} =  \nu 
\pd{x}^{2} v_{0} 
\Eeq

Solving \eqss{leading-order-1}{leading-order-4} in general form requires introducing a potential function  $\varphi(t,z)$. 
\Beq
\label{v-0}
&& v_{0} = S_{0} \pd{z}\varphi \sin(\pi x/\h) \n 
\label{a-0}
&& a_{0} = B\pd{z}\varphi \sin(\pi x/\h)\n
\label{psi-0}
&& \psi_{0} =   \left(\pd{t} \varphi +  \gamma_{1}\varphi\right) \sin(\pi 
x/\h)  \n
\label{b-0}
&& b_{0} =  -\frac{1}{B}\left(f\pd{t} \varphi   + \left( (f
+S_{0})\gamma_{1} - S_{0} \gamma_{2}\right) \varphi \right)\sin(\pi x/\h) 
\Eeq
where
\Beq
\gamma_{1} \ \equiv \ \frac{\pi^{2} \eta}{\h^{2}},\quad 
\gamma_{2} \ \equiv \  \frac{\pi^{2} \nu}{\h^{2}},
\quad
\label{S-0}
S_{0} = - \frac{\pi^{2} B^{2} }{\f \h^{2}}.
\Eeq 

At this point, $\varphi(t,z)$ represents all the dynamical degrees of freedom left undetermined by \eqss{leading-order-1}{leading-order-4}.  This minimalist approach solves \eqss{leading-order-1}{leading-order-4} without adding any more information about the form of the solution until nonlinearity dictates it. This single unknown function represents a scalar-valued ``order parameter'' in the language of critical phenomena.  The typical value of $\varphi(t,z)$ grows in amplitude and becomes increasingly disordered as the system becomes more unstable.  

\Eq{S-0} matches with the critical value 
resulting from linear theory, \eq{S-crit}.  We choose boundary conditions for the magnetic field that match the natural non-dissipative profiles.  Other choice of boundary conditions would bring boundary layers into the analysis. We filter these effects for simplicity.    

The next-order balance produces
\Beq
\label{v1(z)-eq}
&& \pd{z}\left(B v_{1} - S_{0} a_{1}\right) = -J(a_{0},v_{0})  = 0
\n 
\label{a1(z)-eq}
&& \pd{z}\left(B\pd{x}^{2}a_{1} - \f v_{1}\right) = -  J(a_{0},\pd{x}^{2}
a_{0}) = 0 
\n
&&\nonumber \pd{t}a_{1} -B\pd{z}\psi_{1} - \eta \pd{x}^{2}a_{1} = - 
J(\psi_{0},a_{0} ) = \n
&& \label{psi1(z)-eq} \quad \frac{\pi B}{2\h}\sin(2\pi x/\h)\big(  \pd{t}
\left(\pd{z}\varphi \right)^{2}  + 2 \gamma_{1} \left(\pd{z}\varphi 
\right)^{2}  - \pd{z}\left(\left(\pd{t}\varphi + \gamma_{1} \varphi
\right) \pd{z}\varphi \right)\big)\n
&&  \nonumber  \pd{t}v_{1} - B\pd{z}b_{1}  - (\f + S_{0}) \pd{z}
\psi_{1}  - \nu \pd{x}^{2}v_{1}  =   J(a_{0},b_{0}) -  J(\psi_{0},v_{0}) = 
\n
&& \label{b1(z)-eq} \quad -\frac{\pi  (\f-S_{0})}{2 \h}\sin(2\pi x/\h)
\left(  \pd{t}\left(\pd{z}\varphi \right)^{2}  + 2 \gamma_{3} \left(\pd{z}\varphi \right)^{2}  -   \pd{z}\left(\left(\pd{t}\varphi + \gamma_{3} \varphi\right)\pd{z}\varphi 
\right)  \right),
\Eeq
where
\Beq
\gamma_{3} \ \equiv \ \frac{f \gamma_{1} - S_{0} \gamma_{2} }{f -S_{0} }
\Eeq
denotes a weighted average of the magnetic and viscous damping coefficients. 
\Eqs{v1(z)-eq}{a1(z)-eq} determine the strictly $z$-dependent 
components of $v_{1}$ and $a_{1}$, which exactly mimic the form 
of $v_{0}$ and $a_{0}$.  \Eqs{v1(z)-eq}{a1(z)-eq} do not determine 
the $z$-mean components of these variables.  

Averaging \eqs{psi1(z)-eq}{b1(z)-eq} in the vertical direction 
eliminates all pure  $z$ derivatives, and produces the  balances 
\Beq
\label{a1-mean-eq}
&&\left(\pd{t} + 4\gamma_{1}\right)\left< a_{1}\right> \ = \ \frac{\pi B}
{2\h}\sin(2\pi x/\h) \left(\pd{t} + 2\gamma_{1}\right) \left < |\pd{z}
\varphi|^{2} \right>\n
\label{v1-mean-eq}
&&  \left(\pd{t} + 4\gamma_{2}\right) \left< v_{1}\right> \ = \  -\frac{\pi  (\f-S_{0})}{2 \h}\sin(2\pi x/\h)\left(\pd{t} + 
2 \gamma_{3} \right)  \left 
< |\pd{z}\varphi|^{2} \right>,
\Eeq
where  
\Beq
\left< q \right> \equiv \lim_{\H \to \infty} \frac{1}{2\H}\int_{-\H}^{\H} 
q(z)\, \mathrm{d} z.
\Eeq
Defining the amplitudes, 
\Beq
\label{mean A V forms}
\left< a_{1}\right> \equiv \left<A\right> \sin(2\pi x/\h), \quad \left< 
v_{1}\right> \equiv \left<  V\right>  \sin(2\pi x/\h)
\Eeq
simplifies the mean dynamics such that
\Beq
&&\left(\pd{t} + 4\gamma_{1}\right)\left<A\right> = \frac{\pi B}{2\h}
\left(\pd{t} + 2\gamma_{1}\right) \left < |\pd{z}\varphi|^{2} \right>\n
&& \left(\pd{t} + 4\gamma_{2}\right) \left<  V\right>  = -\frac{\pi  (\f-
S_{0})}{2 \h}\left(\pd{t} + 2 \gamma_{3} \right)  \left < |\pd{z}\varphi|^{2} \right>.
\Eeq
The vertical variance of $\pd{z}\varphi$ drives purely $x$-dependent corrections to the mean background field and shear 
momentum.

Closing the system in terms of $\varphi$ requires generating a 
second-order equation for $a_{2}$.  After combining second-order 
versions of \eq{leading-order-1}, 
\Beq
\label{a2-eq}
&&\nonumber B^{3}\pd{z}\left( \pd{x}^{2} + \frac{\pi^{2}}{\h^{2}} 
\right) a_{2} = \n && \quad \frac{\pi  \left(  \left(3 \pi ^2
   B^2+f \h^2 S_{0}\right)\left< 
   A\right>   -B f \h^2
   \left<  V\right> 
   \right)}{\h^3} \pd{z}^{2}\varphi\left(\sin(\pi x/\h) - \sin(3\pi x/ \h)
\right) -\nonumber  \n && \quad \left(\f (\f-S_{0}) \left(\pd{t}^{2} \varphi + 2  \gamma_{3}\pd{t}\varphi +  \gamma_{4} \gamma_{1} \varphi  \right) + 
\f \sigma B^{2} \pd{z}^{2} \varphi + B^{4}\pd{z}^{4} \varphi \right) 
\sin(\pi x/\h),  
\Eeq
where
\Beq
\gamma_{4} \ \equiv \ 2 \gamma_{3} - \gamma_{1}.
\Eeq
The solvability condition requires that all terms multiplying $\sin (\pi 
x/\h)$ on the right-hand side of \eq{a2-eq} cancel identically.  
Otherwise, \eq{a2-eq} admits no finite solutions for $a_{2}$ 
satisfying the boundary conditions \cite{kevorkian_cole}

Grouping all evolution equations together, 
\Beq
\label{WNLMRI-with-dissipation}
\nonumber &&
\left(1 + \Ro \right)\left( \pd{t} + \gamma_{4} \right)\left( \pd{t} +  \gamma_{1}\right) \varphi + \hat{\sigma} \frac{B^{2}}{f} 
\pd{z}^{2} \varphi + \frac{B^{4}}{f^{2}}\pd{z}^{4}\varphi \ =\  0\n
\label{A-evolution}
&&\left(\pd{t} + 4\gamma_{1}\right)\left<A\right> \ =\ \frac{\pi B}{2\h}
\left(\pd{t} + 2\gamma_{1}\right) \left < |\pd{z}\varphi|^{2} \right>\n
\label{V-evolution} && \left(\pd{t} + 4\gamma_{2}\right) \left<  V
\right>  \ =\ -\frac{\pi  f(1 + \Ro)}{2 \h}\left(\pd{t} + 2 \gamma_{3} \right)  \left < |\pd{z}\varphi|
^{2} \right>,
\Eeq
where
\Beq
\Ro \ \equiv \ - \frac{S_{0}}{\f} \ = \ \frac{B^{2} \pi^{2} }{\f^{2} \h^{2}}
\Eeq
defines the ratio of timescales between rotation and critical shear; 
\textit{i.e.}, the Rossby number. Note that $q= 3/4$ corresponds to a Keplerian profile, and $q=1$ corresponds to the Rayleigh instability threshold.  In the case of fixed $q$, we may think of the background magnetic field taking the role of the critical parameter. 

\Eq{WNLMRI-with-dissipation} contains the mean-field shear 
parameter 
\Beq
\label{sigma-hat}
\hat{\sigma} \equiv \sigma -  \frac{\pi  \left(2 \pi ^2 B
   \left<  A\right>  -f \h^2
   \left<  V\right>  \right)}{f
   \h^3}
\Eeq
The feedback from magnetic flux and momentum transport produce 
the only differences between the linear model (including dissipation) 
and a nonlinear model that can saturate the MRI. 

In terms of $q$, 
\Beq
\gamma_{3} \ = \ \frac{\gamma_{1} + q \gamma_{2} }{1+q}, \quad \gamma_{4} \ = \ \frac{(1-q) \gamma_{1} + 2q \gamma_{2} }{1+q}.
\Eeq
For $ 0 < q < 1$, both $\gamma_{3} > 0$ and $\gamma_{4} > 0$ are well-defined averages of  $\gamma_{1}$ and $\gamma_{2}$.  The dissipation coefficients take only one of the two possible orderings 
\Beq
\gamma_{1}  < \gamma_{3} < \gamma_{4} < \gamma_{2}, \quad \mathrm{or}\quad 
\gamma_{2}  < \gamma_{4} < \gamma_{3} < \gamma_{1}.
\Eeq
 
Assuming $\eta = \nu = 0$ allows reducing the system even further. 
In this case, \eqs{A-evolution}{V-evolution} integrate explicitly so 
that
\Beq
\label{ideal mean A, V}
&& \left<A\right> \ = \ \frac{\pi B}{2\h} \left < |\pd{z}\varphi|^{2} \right>
,\quad \left<  V\right>  \ = \ -\frac{\pi  f (1 + \Ro)}
{2 \h}  \left < |\pd{z}\varphi|^{2} \right>
\Eeq

\Beq
\label{WNLMRI}
&&
\left(1+ \Ro \right)\pd{t}^{2} \varphi + \frac{\hat{\sigma}  B^{2} }{f}  
\pd{z}^{2} \varphi + \frac{B^{4}}{f^{2}}\pd{z}^{4}\varphi = 0,
\Eeq
where
\Beq
\label{reduced-sigma-hat}
\hat{\sigma} \ = \ \sigma - \frac{f^3 \Ro (1+ 3 \Ro)}{2 B^2} \left < |
\pd{z}\varphi|^{2} \right>
\Eeq

\Eq{WNLMRI} represents the most important result of this paper.  The next section shows that \eq{WNLMRI} is almost identical to a model of weakly nonlinear loaded elastic beam.  
Holmes \& Marsden found that the forced and damped version of this model possesses an infinite number of chaotic solutions \cite{holmes_marsden}.  This work has been an influential 
paradigm for chaos in PDEs. The remainder of this paper will elaborate on some of the interesting implications for this mathematical analogy between MHD and elastodynamics.  

\Eq{WNLMRI} differs significantly from previous attempts at a weakly nonlinear model for the MRI \cite{umurhan_etal,regev}.  Those cases consider geometry more similar to accretion disks, and as a result required significantly higher (leading-order) dissipation. Past results produce a dissipative Ginzburg-Landau equation, not the high Reynolds number model produced here.  However, for large dissipation \eq{WNLMRI-with-dissipation} reduces to a first-order equation in $\pd{t}$ and hence more like Ginzburg-Landau. 

Liverts and coworkers \cite{liverts_etal} derived an ordinary differential Duffing equation for the thin-disk MRI.  We see in \sec{sec:nonlinear} that \eq{WNLMRI} represents a possible infinity of coupled Duffing equations.  This correspondence shows strong evidence that the ideal Duffing-type dynamics underlies the MRI at a fundamental level. 

\section{Buckling Instability of an Elastic Beam \label{sec:elastic}}

This section derives a dynamical nonlinear model for an elastic 
beam in the vicinity of a buckling instability.  Holmes gave a very short equivocal version of this same derivation \cite{holmes}.  The current context needs enough details to see the similarities and differences to the magnetic case.  We use an asymptotic expansion in terms of the beam aspect ratio.  In this section, 
\Beq
\eps \equiv \frac{\h}{\H} \ \ll \ 1
\Eeq 
where $0 \le z \le \H$ represents the length of the beam along the direction of 
applied load, and $-\h/2 \le x \le \h/2$ represents cross-sectional thickness on the beam.  Figure \ref{diagram}b depicts the basic geometry and background configuration.  A fully complete treatment of fully nonlinear elastodynamics would lead to a very complicated asymptotic analysis in powers of $\eps$. Even though we apply some intuitive reasoning, a completely systematic 
analysis gives the same eventual answer.

The nonlinear strain tensor characterises a general Lagrangian 
deformation of a continuum solid \cite{fung}.  Therefore,
\Beq
| \mathrm{d} \vec{x} + \mathrm{d} \vec{\xi} |^{2} -|\mathrm{d} \vec{x}|
^{2} \ = \  2\, \mathrm{d} \vec{x}\cdot \vec{E}\cdot\mathrm{d} \vec{x} 
\ = \ 2 \left( E_{x,x} \mathrm{d} x^{2} + 2E_{x,z} \mathrm{d} x 
\mathrm{d} z + E_{z,z} \mathrm{d} z^{2}\right)
\Eeq
where $|\mathrm{d} \vec{x}|^{2}  = \mathrm{d} x^{2} + \mathrm{d} 
z^{2}$ represents the \textit{original} distance between nearby 
points, and $| \mathrm{d} \vec{x} + \mathrm{d} \vec{\xi} |^{2}$ 
represents the distance between nearby points after a Lagrangian 
displacement, $\vec{\xi}(t,\vec{x})$.  The explicit components of 
Green's strain tensor are  
\Beq
E_{x,x} = \pd{x}\xi_{x} + \frac{|\pd{x}\xi_{x}|^{2} + |\pd{x}\xi_{z}|^{2}  }
{2}, \quad E_{z,z} = \pd{z}\xi_{z} + \frac{|\pd{z}\xi_{x}|^{2} + |\pd{z}
\xi_{z}|^{2}  }{2}
\Eeq
\Beq
E_{x,z} = E_{z,x} = \frac{1}{2}\left[ \pd{x}\xi_{z} + \pd{z}\xi_{x}+\pd{z} 
\xi_{z} \pd{x}\xi_{z} +\pd{x} \xi_{x} \pd{z}\xi_{x} \right]
\Eeq

We consider a homogenous isotropic Hookean solid with linear stress-strain relationship
\Beq
\vec{S} = \lambda \mathrm{Tr}(\vec{E}) \vec{I}+ 2 \mu \vec{E},
\Eeq
where $\mathrm{Tr}(\vec{E}) = E_{x,x} + E_{z,z}$, and $\vec{I}$ 
represents the identity matrix.  The parameters $\lambda$ and $\mu
$ represent Lamb's constants, but using the 
alternative definitions proves advantageous,
\Beq
 \mu \ \equiv \ \frac{(1-\nu)}{2}Y, \quad
 \lambda  \ \equiv \ \frac{\nu(1-\nu) }{1-2\nu}Y,
\Eeq
where $\nu$ denotes Poisson's ratio (note: this is \textit{not} the 
same as the viscosity parameter in the MRI analysis), and $Y$ 
represents Young's modulus. Some texts use $Y/(1-\nu^2)$ to denote Young's modulus \textit{e.g.}, \cite{fung}.  Our current definition coincides with \cite{holmes}, and allows for a simpler 
derivation.  

The thin aspect ratio leads to small displacements and variation in 
the longitudinal direction, while the perpendicular  displacements 
and variation remain order unity. Kinetic energy also must balance stresses. Therefore, we replace
\Beq
\pd{z} \to \eps \pd{z}, \quad \pd{t} \to \eps^{2} \pd{t}
\Eeq
Considering leading-order balances produces
\Beq
\label{xi_x-eq}
&&\xi_{x} = u(z) - \eps^{2} \left[ \frac{u'(z)^2}{2}  +  \frac{\nu }{1-\nu }
\left( x w'(z) + \frac{x u'(z)^2-x^2
   u''(z)}{2}  \right)\right] + \mathcal{O}(\eps^{4})\n
\label{xi_z-eq} && \xi_{z} = \eps\left(w(z) - x u'(z) \right) + 
\mathcal{O}(\eps^{3}),
\Eeq
where $u(z)$, and $w(z)$ represent unknown mean displacements 
in the $x$ and $z$ directions respectively, and primes, \textit{e.g.}, $u'(z)$, denote z derivatives.  \Eqs{xi_x-eq}{xi_z-eq} imply
\Beq
&& E_{z,z} = \eps^{2} \left(w'(z) + \frac{u'(z)^2}{2} -x u''(z) \right) + 
\mathcal{O}(\eps^{3}), \n
&& E_{x,x} = - \frac{\nu}{1-\nu}E_{z,z} + \mathcal{O}(\eps^{3}), \quad E_{x,z} = \mathcal{O}(\eps^{3}).
\Eeq
These imply the stresses,
\Beq
&&S_{z,z} = \eps^{2} Y \left(w'(z) + \frac{u'(z)^2}{2} -x u''(z) \right) + 
\mathcal{O}(\eps^{3}), \quad S_{x,x},\ S_{x,z} = \mathcal{O}(\eps^{3})
\Eeq
The stress-strain relationship implies the elastic potential energy 
density 
\Beq
\mathcal{U}  \ = \ \frac{1}{2}\sum_{i,j} S_{i,j}E_{i,j} \ = \   
\frac{\eps^{4}}{2} Y \left[w'(z) + \frac{u'(z)^2}{2} -x u''(z) \right]^{2} + 
\mathcal{O}(\eps^{5}),
\Eeq
for which only the $E_{z,z}$ and $S_{z,z}$ components contribute 
to leading order.  
To leading order, only the horizontal displacement contributes to the 
kinetic energy
\Beq
\mathcal{K} = \eps^{4} \rho_{0} \frac{|\pd{t} u|^{2}}{2} + \mathcal{O}
(\eps^{5}).
\Eeq
The explicit $x$ dependence in the potential energy allows 
integrating out this dimension, which produces the effective one-dimensional Lagrangian density
\Beq
&& \mathcal{L} \ \equiv \  \int_{- \h/2}^{\h/2}  (\mathcal{K} - \mathcal{U})\,\mathrm{d}x \ \propto \ \rho_{0} \frac{|\pd{t}u|^{2}}{2} - \frac{Y}
{2} \left[\left( \pd{z}w + \frac{|\pd{z}u|^{2}}{2}\right)^{\!2}  +  
\frac{\h^{2}}{12}|\pd{z}^{2}u|^{2} \right].
\Eeq
The corresponding action follows such that
\Beq
A \equiv \int_{0}^{T} \int_{0}^{\H} \mathcal{L} \dd{z} \dd{t}.
\Eeq
Varying the action with respect to $w(t,z)$ implies  
\Beq
\frac{\partial}{\partial z } \left( \frac{\partial \mathcal{L}}{\partial w'} 
\right) = 0.
\Eeq
Or,
\Beq
\label{Dw-eq}
\pd{z}w + \frac{|\pd{z}u|^{2}}{2}  = \mathcal{E}_{0},
\Eeq
where $\mathcal{E}_{0}$ denotes an integration constant (independent of $z
$, but depending on $t$).
Varying the action with respect to $u(t,z)$ implies that
\Beq
\frac{\partial}{\partial t } \left( \frac{\partial \mathcal{L}}{\partial 
\dot{u}} \right) + \frac{\partial}{\partial z } \left( \frac{\partial 
\mathcal{L}}{\partial u'} \right) = 0
\Eeq
Or,
\Beq
\rho_{0} \pd{t}^{2} u = Y \mathcal{E}_{0}
\pd{z}^{2}u - \frac{Y\h^{2}}{12}\pd{z}^{4} u
\Eeq
Determining $\mathcal{E}_{0}$ in \eq{Dw-eq} requires integrating over the 
entire length of the beam such that
\Beq
\mathcal{E}_{0} = \frac{\Delta \H}{\H} + \frac{1}{2\H}\int_{0}^{\H}|\pd{z}u|^{2} 
\dd{z},
\Eeq
where $\Delta \H \equiv w(z=\H)-w(z=0)$ gives the total change in 
the length of the rod.  

From the definition of Young's modulus applied to the entire solid,
\Beq
\Gamma = - Y \frac{\Delta \H}{\H},
\Eeq
where $\Gamma$ denote the total applied compression ($\Gamma 
> 0$), or tensile ($\Gamma < 0$) load.
Therefore,
\Beq
\pd{z}w =  -\frac{\Gamma}{Y} + \frac{1}{2\H}\int_{0}^{\H}|\pd{z}u|^{2} 
\dd{z} - \frac{|\pd{z}u|^{2}}{2}.
\Eeq
The final reduced dynamical equation for the horizontal deflection 
follows such that
\Beq
\label{BEAM}
\rho_{0}\, \pd{t}^{2} u = \left( -\Gamma  +  \frac{Y}{2\H}\int_{0}^{\H}|
\pd{z}u|^{2} \dd{z} \right) \pd{z}^{2} u - \frac{Y \h^{2}}{12}\pd{z}^{4} u
\Eeq
\Eq{BEAM} is equivalent to the models found in \cite{holmes} and \cite{holmes_marsden}, but with all parameters explicit. 

\section{Discussion \label{sec:discussion}}

\subsection{Dynamical correspondences \label{sec:compare}}

Inspecting each term in \eqs{WNLMRI}{BEAM} highlights the analogies 
between the various physical parameters in the two cases.
To simplify the derivations in \S\ref{sec:MHD}, \eqss{b-eq}{psi-eq} use 
Alfv\'{e}n units such that  $\mu_{0}\rho_{0} = 1$, where $\mu_{0} = 
4\pi$ in cgs units.  This section restores the general parameters in 
order to make comparisons between MHD and elastic parameters.  

The shear criticality in the MHD system and the applied 
compression/tension loading correspond such that,
\Beq
\label{Gamma--sigma}
&& \Gamma \  \longleftrightarrow \ \frac{\sigma }{f + |S_{0}| } 
\frac{B^{2} }{\mu_{0}} \ = \ \frac{\sigma}{f (1+\Ro)} \frac{B^{2}}
{\mu_{0}}
\Eeq
The magnetic pressure, $B^{2}/ \mu_{0}$ lends the correct units to 
the right-hand side of the relationship. 

The magnetic tension in the background magnetic field corresponds 
to Young's modulus in the elastic setting.
\Beq
\label{Y--B^2} && Y \  \longleftrightarrow \   \frac{12 |S_{0}|}{\pi^{2}(f 
+ |S_{0}|)} \frac{B^{2}}{\mu_{0}} \ = \ \frac{12 q }{\pi^{2}(1+q)} 
\frac{B^{2}}{\mu_{0}} 
\Eeq

Together, \eqs{Gamma--sigma}{Y--B^2} relate the shear criticality in 
the fluid to the total strain (relative compression) in the solid.
\Beq
\frac{\Delta L}{L}  \ =\  \frac{\Gamma}{Y}  \  \longleftrightarrow \  
\frac{\pi ^2 \sigma }{12 |S_{0}|} \  = \ \frac{\pi ^2 \sigma }{12 f q}
\Eeq

Lastly, the comparing the nonlinear terms in both systems gives 
interpretation for the scalar potential in MHD in terms of the 
horizontal beam displacement. 
\Beq
\left < |\pd{z}u|^{2} \right>  \  \longleftrightarrow \  \frac{\pi ^4 (f+ 3 | 
S_{0}| )}{12  | S_{0}| \h^2} \left < |\pd{z}\varphi|^{2} \right> \ = \  
\frac{\pi ^4 (1+ 3 \Ro )}{12  \Ro \h^2} \left < |\pd{z}\varphi|^{2} \right>
\Eeq
Or, up to non-dimensional factors,  $u(t,z) \propto \varphi(t,z)/d $.

The potential $\varphi$ does not relate naturally to the 
horizontal Lagrangian displacement in the fluid system.  The definition of the potential function is somewhat arbitrary, but \eqss{v-0}{b-0} produce the most straightforward solution to \eqss{leading-order-1}
{leading-order-4} in terms of derivatives; 
$
b,  \psi   \propto   \pd{t}\varphi; \  a, v  \propto  \pd{z}\varphi.
$
The constants of proportionality are more chosen according to style.  
In terms of (fluid) Lagrangian displacements, 
$ \xi_{x} \propto \pd{z} \varphi; \ \xi_{z} \propto  \varphi $. 
Whereas for the elastic solid
$\xi_{x} \propto u, \pd{z}w; \ \xi_{z} \propto  w,  \pd{z} u$.
The correspondence between the parameter in the MRI and the 
buckling beam would allow demonstrating some 
aspects of MHD with a more manageable slender elastic rod. 

\subsection{Saturation \label{sec:saturation}}

How the MRI saturates remains an open question in the 
context of accretion disk modelling \cite{julien_knobloch,liverts_etal}.  Intuitive understanding in the case of the elastic beam, helps clarify the saturation mechanism of the MRI in this simple model. An elastic beam saturates a buckling instability by setting one side of the beam under tension and the other side under compression. In a sense, the instability transports stress (equivalently strain and density) 
dynamically from the tension side to the compression side.  In the MHD case, \eq{mean A V forms} and \eq{ideal mean A, V} imply that the instability puts the left-half ($0 < x < \h/2$) of the domain under a sleight excess of magnetic flux and sleight deficit of linear momentum. The right-half ($\h/2 < x < \h$) receives the opposite feedback such that the total of both quantities remains conserved.  In the linear phase of the instability, momentum and magnetic flux compete to both drive and suppress growing perturbations.  The nonlinear transport rearranges the background so that both halves of the domain experience more stringent stability criteria.  In their early MRI studies, Balbus and Hawley demonstrated positive outward angular momentum flux \cite{H&B92}. From \eq{ideal mean A, V}, $\left<V\right> < 0$  also corresponds to positive outward transport. Even in the local, incompressible, Cartesian geometry, the system can correctly discern inward and outward.  Along with angular momentum transport, the exchange of magnetic potential reconciles with past accretion disk models \cite{ebrahimi_etal}.

\subsection{Symmetry \label{sec:symmetry}}

The buckling beam analogy is not the first between MHD and elastic systems \cite{ogilvie_proctor,ogilvie_potter}.   There are deep mathematical reasons for this relationship. Chandrasekhar speculated that the reason for not recovering Rayleigh's criteria \eq{Rayleigh-condition} in the limit of zero magnetic field ``must lie in the circumstances that ... the lines of magnetic force are permanently attached to the fluid" \cite{chandra_b}.  This is another way of saying that specifying the location of the magnetic field lines is the same as specifying the location of the fluid parcels.  

Like an elastic medium, MHD supports shear stresses and shear waves in the form of Alfv\'{e}n displacements. Symmetry breaking provides the mechanism underlying this simple physical fact.
When considering a collection of $N$ particles, phase space generically comprises $3N$  momenta and $3N$ coordinates for a total of $6N$ dimensions. For a fluid, the range scales of fluctuations determines the effective $N$.  Independence of the system on one or more of the coordinates implies conservation of the corresponding momenta.  Ordinary fully compressible hydrodynamics is only $5N$ dimensional, $3N$ for the velocities, $1N$ for the density, and $1N$ for entropy (or equivalently the pressure). The missing third set of coordinates implies the conservation of potential vorticity (effectively a component of momenta).  There is no restoring force for one component of the momentum (more accurately one subset of size $N$).  For incompressible hydrodynamics a much smaller set of unique coordinate labels implies Kelvin's Circulation Theorem which implies an much larger set of conserved momenta. Magnetic field changes this picture.

The frozen-in condition \cite{AT42} provides the magnetic fluid with enough Lagrangian coordinate labels to break potential vorticity conservation.  In fact, MHD seems to provide more coordinate labels than necessary to specify a location in phase space completely.  The divergence condition, and other constraints, remedy the over-counting problem; see \cite{vasil_etal} for more details.  With a full accounting for fluid parcels, there remain no point-wise conserved momenta, and almost all degrees of freedom must experience a non-zero force of some kind.  Lack of Lagrangian-particle relabelling symmetry is an imperative feature MHD shares with elastodynamics.  In the latter case, the dynamics depends explicitly on all components of the displacement field through the strain tensor.    

Along with their primary discovery, Balbus and Hawley produced an analogy between magnetic field lines between Keplerian fluid parcels and a simple elastic spring connecting two orbiting masses \cite{B&H92}.  In this example, the connecting spring breaks the conservation of the individual masses angular momentum and Laplace-Runge-Lenz vectors. This example shows that the tension in the magnetic field and/or spring provides both stabilisation via tension, and acts to liberate rotational energy.  The buckling beam differs in this respect.  The product of magnetic field and shear compares to the compression in \eq{Gamma--sigma}. The magnetic field alone (without destabilising shear) compares to Young's modulus in \eq{Y--B^2}.

\subsection{Nonlocality \label{sec:nonlocality}}

The nonlocal nature of the nonlinear feedback requires some consideration.  The nonlinear feedback in \eq{WNLMRI} appears surprising at first. Especially given that the original MHD model does not seem to contain nonlocal terms.  How does the non-locality arise? 

Answering this requires pointing out a number of small facts. \textit{(I):} In spite of appearances the original equations \eqss{b-eq}{psi-eq} are actually nonlocal.  This results from evolving $\omega_{y} = \nabla^{2} \psi$, rather than $\psi$ alone in \eq{psi-eq}. Computing the inverse Laplacian amounts to convolving against the appropriate Green's function. This is a non-local operation.  \textit{(II):} Unlike the incompressible MHD equations, the full elastodynamic equations are completely local.  \textit{(III):} Therefore, the reason for non-locality in \eq{WNLMRI} cannot result from the non-locality \eq{psi-eq}. But similar physics underlies both situations.  \textit{(IV):} Non-locality in incompressible MHD results physically from fast acoustic waves equalising the pressure field on timescales much faster than dynamical timescales. Rather than a experiencing a small delay, distant points respond instantaneously to fluid motions, but in a fashion that attenuates with distance away from the disturbance.  \textit{(V):} The buckling beam achieves non-locality through of the separation of timescales between extremely fast elastic waves, and unstable flexural waves.  \textit{(V):} In addition to the unstable MRI modes, the full linear dispersion relation \eq{full dispersion} contains two additional fast modes for each wavenumber. \textit{(VI):} These mixed Alfv\'{e}n-Coriolis modes propagate with the phase speed
\Beq
\frac{\omega_{\textit{fast}}}{k} \ \approx \ \pm \sqrt{B^2 + \frac{\f^{2}\h^2}{\pi^{2}}} + \mathcal{O}(\sigma^{2}).
\Eeq
\textit{(VII):}
Both magnetic tension and the Taylor-Proudman effect \cite{TP23} provide rigidity in the vertical direction and transmit magnetic and kinetic stresses through the fluid. 

Incompressibility implies that non-locality is more common than often realised.  The surprising aspect about \eq{WNLMRI} is that feedback is entirely non-local.  \eq{reduced-sigma-hat} shows no spatial variation or attenuation. We could express \eq{reduced-sigma-hat} in local form by stating $\pd{z} \hat{\sigma} = 0$; but we require some other condition to determine the actual value of $\hat{\sigma}$.  Conservation of total-integrated momentum and magnetic flux provides the solution.  The global conservation of these quantities also underlies the saturation mechanism.

\subsection{Dissipative stability \label{sec:dissipation}}

Also, in contrast to the ideal stability theory in \S\ref{sec:MHD}\ref{linear-theory} the nonlinear analysis elucidates stability with a small amount of diffusion.  In that case there exist a critical shear curve as a function of wavenumber, \textit{i.e.},
\Beq
\label{dissipation-sigma(k)}
\sigma_{c}(k) =  \frac{B^2
   k^2}{f} + \frac{(1+q) \gamma_1 \gamma_4 f
   }{B^2 k^2},
\Eeq 
where the first term on the right-hand side gives the stability curve in the ideal limit. Minimising over $k$, gives
\Beq
k_{c} = \frac{(1+q)^{1/4} \gamma_1^{1/4} \gamma_4^{1/4} \sqrt{f}
   }{B},\quad \sigma_{c}(k_{c}) = 2 \sqrt{(1+q)\gamma_1 \gamma_4}.
\Eeq
Figure \ref{stability regions}b shows the difference in stability for dissipative versus non-dissipative dynamics. This adds further insight into breaking the apparent degeneracies associated with the small-dissipation limit of the MRI; also see \cite{kirillov_etal_a,kirillov_etal_b}. In the limit as the background magnetic field vanishes, we find a necessary condition in terms of the magnetic Reynolds number,
\Beq
\label{Re_M,c-eq}
\textit{Re}_{\textit{M}} \equiv - \frac{S\h^{2}}{\eta} > 2 \pi^{2} + \mathcal{O}(B^{2}).
\Eeq

\section{Nonlinear solutions \label{sec:nonlinear}}

Rescaling the time, space and amplitude variables simplifies the 
discussion of nonlinear solutions to \eqs{WNLMRI}{BEAM}.  
Assuming $\eta = \nu = 0$, and defining a non-dimensional length, time, and amplitude leads to
\Beq
\label{scaled-wnlmri}
\pd{t}^{2} \varphi + \left(\mu - \left< |\pd{z}\varphi|^{2} \right>\right)
\pd{z}^{2}\varphi + \pd{z}^{4} \varphi = 0,
\Eeq
where $\mu = \pm 1$. It is straight forward in both the MHD and elastic cases to find the relevant re-scalings that yield \eq{scaled-wnlmri}. 


The completely nonlocal character of the nonlinearity in \eq{scaled-wnlmri} implies a particularly interesting class of solutions.  The spatial Fourier transform  
\Beq
\varphi_{k}(t) \ \equiv \ \frac{1}{\sqrt{2\pi}}\int_{-\infty}^{\infty} \varphi(t,z)e^{-i k z } \, 
\mathrm{d}z
\Eeq
implies 
\Beq
\label{Fourier-wnlmri}
\ddot{\varphi}_{k}(t) \ = \ k^{2}\left(\mu - k^{2} - \int_{-\infty}^{\infty}
k^{2} |\varphi_{k}(t)|^{2} \,\mathrm{d}k\right)\varphi_{k}(t)
\Eeq
Assuming $\varphi(t,y)$ is real implies the Hermitian symmetry $
\varphi_{k}^{*}(t) = \varphi_{-k}(t)$. Each mode linearly self interacts, 
and couples to all other active modes only through the integrated 
feedback.  For periodic solutions, the integral in \eq{Fourier-wnlmri} 
becomes a discrete sum. It is now clear that system is equivalent to a large collection of mean-field interacting Duffing oscillators. 

\Bfig
\begin{center}
\includegraphics[scale=0.17]{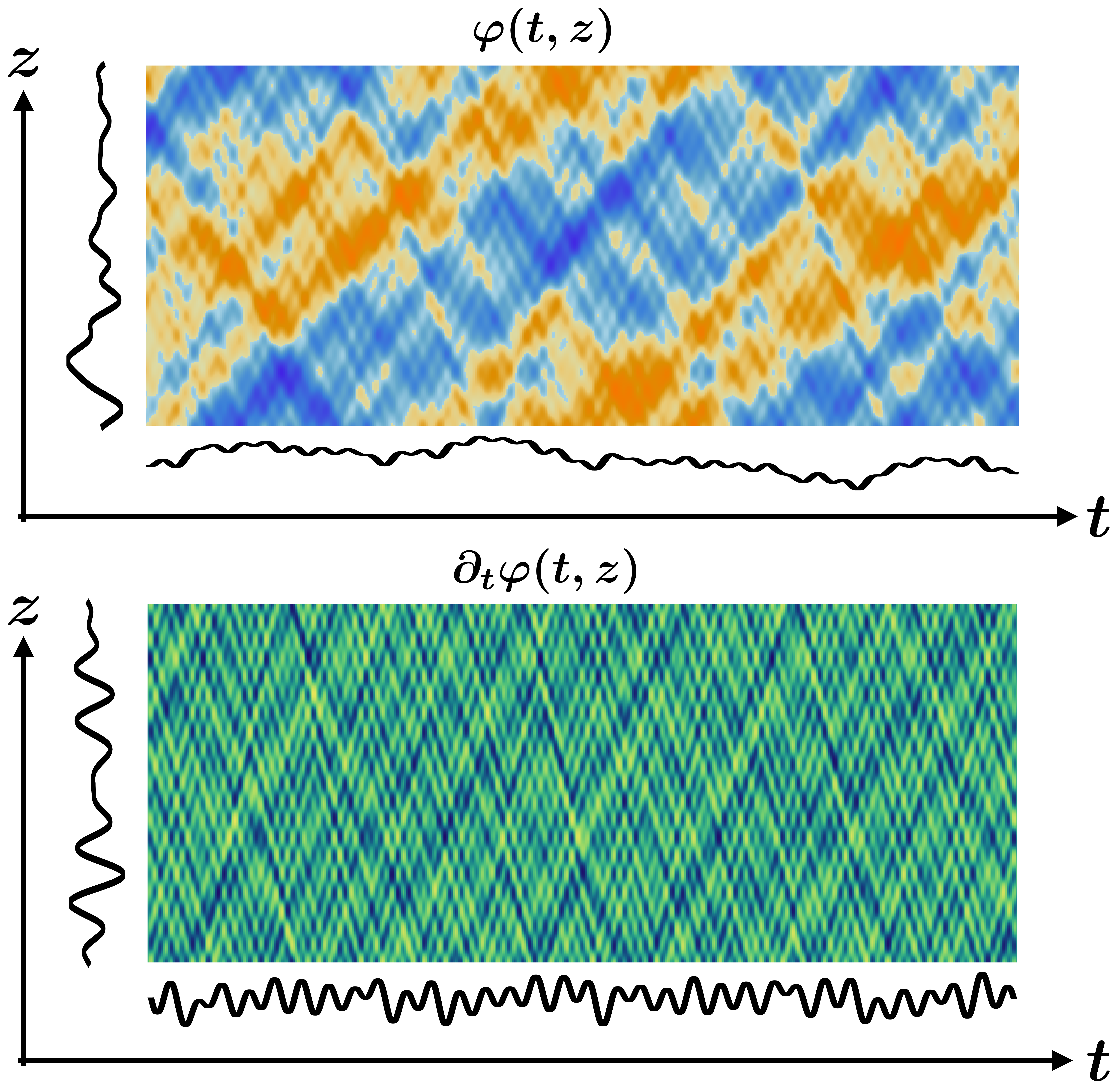}
\caption{Solutions to \eq{scaled-wnlmri} for random initial 
conditions.  
The vertical squiggle on the left-hand side of each image represents 
the initial state of $\varphi(t=0,z)$ and $\pd{t}\varphi(t=0,z)$ 
respectively. The horizontal squiggle at the bottom of each image 
represents a slice through each space-time diagram at $z=1/2$. 
The $z$-axis runs from $0$ to $2\pi$, and time runs from $0$ to  
$10$. For $\varphi$, the plots are scaled between $\pm 1.5$. For $
\pd{t} \varphi$, the plots are scaled between $\pm 4$. 
\label{space-time-diagram}}
\end{center}
\Efig

The original nonlinear system (with $\eta = \nu = 0$) is conservative, and the reduced system remains Hamiltonian with the total conserved energy 
\Beq
\label{Energy}
&&  H \equiv \frac{1}{2}\int_{-\infty}^{\infty}\left( |
\dot{\varphi}_{k}(t)|^{2} - k^{2}(\mu - k^{2})|\varphi_{k}(t)|^{2}  \right)
\mathrm{d}k   +\frac{1}{4}\left(\int_{-\infty}^{\infty} 
k^{2} |\varphi_{k}(t)|^{2}  \mathrm{d}k \right)^{2}.
\Eeq
\eq{Fourier-wnlmri} results canonically from \eq{Energy}. 

The completely mean-field character of \eq{Fourier-wnlmri} implies 
interesting symmetries when considering the evolution of the complex-valued amplitude in terms of the real amplitude and phase 
\Beq
&&\varphi_{k}(t) \ = \ r_{k}(t)e^{i \theta_{k}(t)}.
\Eeq
In in this case, the feedback is independent of the phase such that 
the imaginary component of \eq{Fourier-wnlmri} implies
\Beq
\frac{d}{dt}\left( r_{k}(t)^{2} \dot{\theta}_{k}(t) \right) = 0.
\Eeq
That is, \textsl{for each} individual value of $k$, each ``angular 
momentum" remains constant in time
\Beq
L_{k} \equiv r_{k}(t)^{2} \dot{\theta}_{k}(t)
\Eeq
The conservation of $L_{k}$ results from the symmetry for each $k$, $\varphi_{k} \to \varphi_{k} \exp(i \chi_{k})$, for any time-independent $\chi_{k}$.

The system further reduces to the following real equations 
\Beq
\label{real-wnlmri-amp} \ddot{r}_{k}(t) - \frac{L_{k}^{2}}{r_{k}
(t)^{3}}= k^{2}\left(\mu - k^{2} -\int_{-\infty}^{\infty}k^{2} r_{k}(t)^{2} \,
\mathrm{d}k\right)r_{k}(t),\quad \dot{\theta}_{k}(t) \ = \ \frac{L_{k}}{r_{k}(t)^{2}}
\Eeq
The angular momentum $L_{k}$ is an arbitrary function of $k$, and 
is independent of time. The total energy now becomes 
\Beq
H = \frac{1}{2}\int_{-\infty}^{\infty} \left(\dot{r}_{k}(t)^{2}  + 
\frac{L_{k}^{2}}{r_{k}(t)^{2}} - k^{2}(\mu-k^{2})r_{k}(t)^{2}\right)
\mathrm{d}k +  \frac{1}{4}\left[\int_{-\infty}^{\infty}k^{2}r_{k}(t)^{2}\,\mathrm{d}k 
\right]^{2}
\Eeq

In terms of the complex initial data,
\Beq
\varphi_{0,k} = \varphi_{k}|_{t=0}, \quad \dot{\varphi}_{0,k} = 
\dot{\varphi}_{k}|_{t=0},
\Eeq
time-integrating \eq{real-wnlmri-amp}, 
requires the angular momenta, phases, radii, and radii velocities. 
Respectively, 
\Beq
&& L_{k} \ = \ \mathrm{Im}\!\left[\varphi^{*}_{0,k}\dot{\varphi}_{0,k}  
\right],\quad
\theta_{k}(0) \ = \ \mathrm{Im}\!\left[\log \varphi_{0,k} \right] \n
&& r_{k}(0) \ = |\varphi_{0,k}|, \quad \dot{r}_{k}(0) \ = \ 
\frac{\mathrm{Re}\!\left[\varphi^{*}_{0,k}\dot{\varphi}_{0,k}  \right]}{|
\varphi_{0,k}|}
\Eeq

Furthermore, the inverse Fourier transform of $L_{k}$ gives a point-wise conserved quantity in the spatial domain, 
\Beq
\Lambda(t,z) \equiv \int\left[ \,\dot{\varphi}(t,z + z')\varphi(t,z') - 
\varphi(t,z + z')\dot{\varphi}(t,z')\, \right]\mathrm{d}z',
\Eeq
which represents the anti-symmetric component of the cross-correlation function. For every $z$, $\pd{t}\Lambda(t,z) = 0$. The conservation of $\Lambda(t,z)$ is equivalent to the conservation of 
\Beq
\Lambda_{s}(t) \equiv 
\int  
\pd{t}\varphi(t,z) \pd{z}^{s} \varphi(t,z)\mathrm{d}z,
\Eeq
for all odd values of $s$. The case $s=1,3$ respectively correspond to magnetic-helicity and cross-helicity  conservation in the original MHD system. Even thought the system contains an infinite number of conserved quantities, it is not apparently integrable. 

\eq{real-wnlmri-amp} shows that the dynamics is equivalent to a 
collections of particles all moving in an averaged potential, and 
each individually conserving its angular momentum.  The nonlocal character of the nonlinearity implies that \textsl{every} Fourier truncation of the system represent and \textsl{exact} solution.  In other words, the system does not cascade energy to smaller scales than are present in the initial conditions. This unusual form of memory implies an extremely diverse class of solutions.  The model displays one of the simplest types of mean-field dynamics, but with non-trivial results.  

Figure~\ref{space-time-diagram} shows a typical solution for a modest number of initial modes.  The solution lies in a periodic domain $0 \le z \le 2 \pi$, with $\mu = +1$. The initial conditions consist of $K=20$ Gaussian-random unit-amplitude complex-valued Fourier modes for both $\varphi$ and $\dot{\varphi}$. Using a 4th-order adaptive Runge-Kutta scheme, we time integrate the system of ODE's for 30 non-dimensional time units,   
\Beq
\label{discrete-Fourier-wnlmri}
\ddot{\varphi}_{k}(t) \ = \ k^{2}\left(\mu - k^{2} - \sum_{k=1}^{K}
k^{2} |\varphi_{k}(t)|^{2} \right)\varphi_{k}(t)
\Eeq
The spatial form of the solution is reconstructed such that
\Beq
\varphi(t,z) = \sum_{k=1}^{K} \varphi_{k}(t) \exp{(i k z)}  + \mathrm{c.c.}
\Eeq
The computations were computed with the Dedalus code.  Dedalus is a very general toolkit for computing the pseudo-spectral solution to a large number of PDE's and related problems. For more information and links to the source code, see \url{dedalus-project.org}. 

Figure~\ref{space-time-diagram} shows a pattern of large-scale traveling waves with faster small-scale dynamics superimposed.  The small-scale fast dynamics show up clearly in the velocity pattern, $\pd{t}\varphi$.
The qualitative aspects of this pattern seem typical for an array of initial conditions.  A more comprehensive numerical study of the space of solutions both with and without dissipation lies beyond the scope of this paper. For an example of the possible range rich behaviours, Eugeni and coworkers recently studied the forced nonlinear beam dynamics with multiple interacting degrees of freedom \cite{eugeni_etal}.

\section{Conclusions \label{sec:conclude}}

Heuristically, one can use the concept of entropy to interpret two 
different physical systems producing the same reduced governing 
equations. Near an instability, only a small number of degrees of 
freedom grow to significance.  There are many fewer ways to 
organise a small number of modes than the large number active in 
a strongly driven system.  Or, to borrow from Tolstoy, "Weakly 
nonlinear systems are all alike; every strongly nonlinear system is 
nonlinear in its own way."\footnote{Tolstoy's original quote: "Happy 
families are all alike; every unhappy family is unhappy in its own 
way.", concisely describes how some macro states can contain vastly different numbers of possible micro states.  Anna Karenina was published within a couple years of Boltzmann's original mathematical formulation of entropy.}  This is especially true in the presence of symmetry.  This fact 
is apparent in that so many systems reduced to a small set of well-known canonical equations. For example, the complex Ginzburg-Landau 
equation (and its further simplifications) derives systematically from 
an extremely wide range of physical starting points \cite{cross_hohenberg}.  

This paper points out a new class of \textit{nonlocal} 
models deriving from at least two different natural systems. More physical examples likely exist. In a particularly fundamental way, MHD is more similar to elastodynamics than pure hydrodynamics.  The magnetic field provides a final material coordinate label, and therefore eliminates fluid particle relabelling symmetry, and conservation of potential vorticity.  In elastodynamics, material particles contain explicit labels, and the issue of potential vorticity never really arises.  This implies that fluid systems without conserved potential vorticity, and a spring-like restoring mechanism could produce similar dynamics as the MRI or buckling beam. Over-stable double-diffusive convection satisfies these conditions, and (speculatively) may provide an additional example. It is quite possible that this mean-field network of Duffing oscillators forms a new class of universal near-equilibrium dynamics.   

More generally, there are infinitely more ways to 
produce a nonlocal interaction than a local one. The multiple scales assumption allows for information to travel asymptotically fast and produce the instantaneous interaction 
we see in our model.  In both the buckling beam, and MRI, the reason for the particular 
type of nonlinearity arises from conservation laws.  In the elastic case, the 
integral terms result from fixing the total mass of the beam. In the 
MHD case, the integral terms arise from conservation of total linear 
momentum, and magnetic flux.  In both cases, these quantities 
remain conserved with the inclusion of dissipative effects.  Rotating and/or magnetised free-slip Rayleigh-B\'{e}nard convection, displays nonlocal nonlinear terms arising from conservation of total momentum and/or magnetic flux; just as in the MRI problem \cite{cox_matthews,beaume_etal}.  These correspondences perhaps imply that global conservation principles and non-locality relate more deeply.  Intriguingly, the nonlocal terms in the MRI model allow for a very large class of symmetries and this hints at the deeper link to conservation principles.

Elastodynamics and MHD likely part ways with the introduction 
of more physical ingredients.  The systems are likely very different from each other in three dimensions.  Dissipative boundary layer dynamics (filtered in this paper with judicious choice of boundary conditions), interactions with other fluid instabilities, and more complex 
geometric effects all likely pull the correspondence further apart.  Nevertheless, the link between elastic buckling and the MRI produces interesting insight regarding the growth and saturation of the MRI, albeit in a regime far removed from traditional astrophysical applications.  The type of derivation presented in this paper would likely work in a thin disk-like geometry, which is more suitable to accretion disks. This interesting case will produce several additional complications resulting from boundary conditions alone.  Attempting to understand more complex systems requires adding dynamical elements  back into the minimalist model and considering the consequences.  Hopefully adding richer dynamical ingredients will still allow simple mathematical progress.  But even when interactions become too numerous to consider analytically,  understanding the relevant spatiotemporal scales and possible couplings will greatly help streamline numerical simulation of more complex models.

\medskip

\noindent \textbf{Competing interests statement:} I have no competing interests
 
\medskip
 
\noindent \textbf{Data accessibility statement:} This work does not have any experimental data.

\medskip

\noindent \textbf{Funding:} The Australian Research Council supports the author through a Discovery Early Career Researcher Award, number DE140101960. Part of this work was conducted with funding from the University of California Berkeley Theoretical Astrophysics Center. 

\medskip

\noindent \textbf{Acknowledgements:} The author would like to thank Edgar Knobloch for originally pointing out Holmes and Marsden's work on elastic buckling. The computations used the Dedalus code: \url{dedalus-project.org}.  The author thanks the remainder of the Dedalus collaboration: Keaton Burns, Daniel Lecoanet, Jeff Oishi, and Ben Brown for their significant contributions to the code project.  The author thanks two anonymous referees for suggesting improvements to the manuscript.

\end{document}